\documentclass[preprint,showpacs,prd,floatfix,nofootinbib,tightenlines,11pt]{revtex4-1}
\usepackage{graphicx}
\usepackage[dvips,usenames]{color}
\usepackage{amsmath,amssymb}
\newcommand{\bs}{\begin{sloppypar}} \newcommand{\es}{\end{sloppypar}}
\def\beq{\begin{eqnarray}} \def\eeq{\end{eqnarray}}
\def\beqstar{\begin{eqnarray*}} \def\eeqstar{\end{eqnarray*}}
\newcommand{\bal}{\begin{align}}
\newcommand{\eal}{\end{align}}
\newcommand{\beqe}{\begin{equation}} \newcommand{\eeqe}{\end{equation}}
\newcommand{\p}[1]{(\ref{#1})}

\begin {document}
\title{Anisotropic pressure in strange quark matter \\ under the presence of a strong magnetic field
 }
\author{ A. A. Isayev}
\email{isayev@kipt.kharkov.ua}
 \affiliation{Kharkov Institute of
Physics and Technology, Academicheskaya Street 1,
 Kharkov, 61108, Ukraine
\\
Kharkov National University, Svobody Sq., 4, Kharkov, 61077, Ukraine
 }
  \author{J. Yang}
 \email{jyang@ewha.ac.kr}
 \affiliation{Department  of Physics and the Institute for the Early Universe,
 \\
Ewha Womans University, Seoul 120-750, Korea
}
\begin{abstract}
Thermodynamic properties of strange quark matter in strong magnetic
fields $H$ up to $10^{20}$~G are considered within the MIT bag model
at zero temperature
  implying the constraints of total baryon number
conservation, charge neutrality and chemical equilibrium. The
pressure anisotropy, exhibiting in the difference between the
pressures along and perpendicular to the field direction, becomes
essential at $H>H_{th}$,  with the estimate
$10^{17}<H_{th}\lesssim10^{18}$~G. The longitudinal pressure
vanishes in the critical field $H_c$, which can be somewhat less or
larger than  $10^{18}$~G, depending on the total baryon number
density and bag pressure. As a result, the longitudinal instability
occurs in strange quark matter, which precludes: (1)~a significant
drop in the content of $s$  quarks, which, otherwise,
 could happen at $H\sim10^{20}$~G; (2) the
 appearance
 of positrons in weak processes in a narrow interval
near $H\sim2\cdot10^{19}$~G (replacing electrons). The
 occurrence of the longitudinal instability  leaves the possibility only for
electrons  to reach a fully polarized state, while for all quark
flavors the polarization remains  mild even for the fields near
$H_c$. The anisotropic equation of state   is determined under the
conditions relevant to the interiors of magnetars.
\end{abstract}
  \maketitle

Strange quark matter, composed of deconfined $u,d$ and $s$ quarks,
can be the true ground state of matter, as was suggested in
Refs.~\cite{AB,W,FJ}. There it was found that, at zero temperature
and pressure, the energy per baryon in strange quark matter for a
certain range of the model QCD-related parameters can be less than
that for the most stable $^{56}$Fe nucleus. This conjecture, if will
be confirmed, would have important astrophysical implications. In
particular, strange quark matter can form strange quark stars
self-bound by strong interactions~\cite{I,AFO,HZS}. This is in
contrast to an ordinary scenario, in which  neutron stars are
composed of hadrons (plus some admixture of leptons to ensure charge
neutrality and beta equilibrium) and are bound by gravitational
forces. Also, if strange quark matter is metastable at zero
pressure, it can appear in the high-density core of a neutron star
as a result of the deconfinement phase transition. In this case, the
stability of strange quark matter is provided by the gravitational
pressure from the outer hadronic layers. Then a relevant
astrophysical object is a hybrid star having a quark core and the
crust of hadronic matter. Moreover, strange quark matter can be
potentially encountered in the form of small nuggets called
"strangelets"~\cite{FJ,WFLP}. 
For the review on the properties of strange quark matter and its
possible forms in astrophysics, one can address to
Refs.~\cite{FW,M}.

Another important aspect related to the physics of compact stars is
that they are endowed with the magnetic field. In particular, for
 conventional neutron stars the  magnetic
field strength at the surface can reach the values in the range of
$10^9$-$10^{13}$~G~\cite{LGS}. For the special classes of neutron
stars such as soft $\gamma$-ray repeaters (SGRs) and anomalous
$X$-ray pulsars (AXPs),  the field strength can reach even larger
values of about $10^{14}$-$10^{15}$~G~\cite{TD,IShS}. These strongly
magnetized objects are called magnetars~\cite{DT}. It was also
suggested that a  magnetized hybrid star, or a magnetized strange
quark star can be a real source of the SGRs or AXPs~\cite{CD,OLN}.
The actual mechanism, by which magnetars generate such strong
magnetic fields, is still under debate. Together with the dynamo
amplification scenario in a magnetar  with the fast rotating
core~\cite{TD}, other possibilities such as spontaneous ordering of
hadron~\cite{IY}, or quark~\cite{TT} spins are not excluded.

Under such circumstances, the issue of interest is the impact of a
strong magnetic field on thermodynamic properties of neutron star
matter~\cite{CBP,BPL,CPL,PG,IY4,IY10,IY11}, or hybrid/quark star
matter~\cite{C,BPC,RPPP,WSYP,FMRO}. Note that in the interior of a
magnetar magnetic field strength can reach even larger values
compared to that at the surface. In the recent study~\cite{FIKPS},
it was shown that, for either of the  scenarios with a
gravitationally bound neutron star or with a self-bound quark star,
the field strength in the magnetar core could be as large as
 $10^{20}$~G. In such ultrastrong magnetic fields, the effects of
 the $O(3)$
 rotational  symmetry breaking by the magnetic field become
 important~\cite{FIKPS,Kh,IY_PLB12,IY_PRC11}.  In
 particular, the longitudinal (along the magnetic field) pressure is
 less than the transverse (perpendicular to the magnetic field)
 pressure resulting in the appearance of the longitudinal
 instability of the star's matter if the magnetic field exceeds some
 critical value. The effects of
 the pressure anisotropy should be accounted for in the consistent
 study of structural and polarization properties of  a strongly magnetized
 stellar object. As a consequence of these effects,
 the equation of state (EoS) of the stellar matter becomes essentially
 anisotropic in an ultrastrong magnetic field.

In the given research, we  consider the effects of the pressure
anisotropy in strange quark matter under the presence of a strong
magnetic field. To study quark matter, we use a phenomenological MIT
bag model~\cite{CJJ}, in which quarks of various flavors are
considered as degenerate Fermi gases, confined in a finite region of
space by the inward bag pressure. The MIT bag model provides simple
and physically transparent framework to approach the problem and was
applied earlier to the study of nonmagnetized~\cite{FJ,AFO,HZS} and
magnetized~\cite{C,BPC,RPPP,WSYP,FMRO} strange quark matter. Note
that the effects of the pressure anisotropy  were not included in
the research of magnetized strange quark matter  in
Refs.~\cite{C,BPC,RPPP,WSYP}. Besides, as we will discuss in more
detail further, the determination of the total anisotropic pressure
in Ref.~\cite{FMRO} was incomplete, where, in fact, the pure field
contribution (the Maxwell term)  was missed. However, just this term
provides the most principal source of the pressure anisotropy in
strong magnetic fields $H>H_{th}$,  where
$10^{17}<H_{th}\lesssim10^{18}$~G ($H_{th}$ is the threshold field
at which the pressure anisotropy becomes relevant). As a result, the
longitudinal $p_l$ and transverse $p_t$ pressures interchange their
roles: instead of inequality $p_t<p_l$ obtained in Ref.~\cite{FMRO},
one gets the opposite, $p_t>p_l$.

It is worthy to note at this point, that at sufficiently high
density strange quark matter will be in color superconducting
color-flavor-locked (CFL) state~\cite{ARW,LH}, in which quarks of
all flavors and colors near the Fermi surface are paired.
Nevertheless, it is unknown which of the phases, color
superconducting or normal, will be preferable at the densities
closer to the density of  the deconfinement phase transition. Recent
perturbative QCD calculations up to the second order on the coupling
constant $\alpha_s$ with the finite strange quark mass still do not
allow nor confirm nor rule out the existence of unpaired strange
quark matter, if to take into account the uncertainties in the model
parameters~\cite{KRV}. By this reason, the study of the impact of a
strong magnetic field on the thermodynamic properties of normal
strange quark matter is expedient.

\section{General formalism}
In the simplest version of the MIT bag model, quarks are considered
as free fermions moving inside a finite region of space called a
"bag".  The effects of the confinement are accomplished  by endowing
the finite region with a constant energy per unit volume, the bag
constant $B$.  The bag constant $B$ can be also interpreted as the
inward pressure - the "bag pressure", needed to confine quarks
inside the bag. In the MIT bag model, the bag constant is considered
as a phenomenological parameter of the theory. Therefore, all
relevant equations can be obtained, first, by considering the
relativistic degenerate system of free fermions in an external
magnetic field, and then, in order to get the equation of state of
the system, by properly modifying the energy and the pressure of
magnetized noninteracting fermions  with account of the  bag
constant.

The Lagrangian density for the relativistic system of
 noninteracting quarks ($u, d$ and $s$) and leptons ($e$) in an external magnetic
 field reads
 \begin{equation}\label{L}
    {\cal L}=\sum_{i=u,d,s,e} \bar\psi_i[\gamma^\mu(i\partial_\mu-q_i 
A_\mu)-m_i]\psi_i-\frac{1}{16\pi}F^{\mu\nu}F_{\mu\nu}. 
\end{equation}

Further we will assume that the external magnetic field is directed
along the $z$ axis, and, correspondingly, choose the $4$-potential
in the form $A_\mu=Hx_1\delta_{\mu2}\; (\mu=0,1,2,3)$, where $H$ is
the magnetic field strength.  In Eq.~\p{L}, $F_{\mu\nu}=\partial_\mu
A_\nu-\partial_\nu A_\mu$ is the electromagnetic field tensor.
 Analogously to
Refs.~\cite{BPC,RPPP,WSYP}, we did not take into account
in~Eq.~\p{L} the quark anomalous magnetic moments which are not
completely understood in the deconfined state. Also, we disregarded
the electron anomalous magnetic moment, whose role, as believed, is
insignificant even for the magnetic fields encountered in
magnetars~\cite{D}. The quark $\psi_f$ ($f=u,d,s$) and electron
$\psi_e$ spinors satisfy the Dirac equation
\begin{align}[\gamma^\mu(i\partial_\mu-q_i 
A_\mu)-m_i]\psi_i=0, \quad i=u,d,s,e.\end{align}

The energy spectrum of free relativistic fermions in an external
magnetic field has the form~\cite{BPL}
\begin{align}\varepsilon^i_{\nu}=\sqrt{k_z^2+m_i^2+2\nu |q_i|H}, \quad
\nu=n+\frac{1}{2}-\frac{s}{2}\,
\mathrm{sgn}(q_i),\label{spectr}\end{align} where $\nu=0,1,2...$
enumerates the Landau levels,  $n$ is the principal quantum number,
$s=+1$ corresponds to a fermion with  spin up, and $s=-1$ to a
fermion with  spin down. The lowest Landau level with $\nu=0$ is
single degenerate and other levels with $\nu>0$ are double
degenerate. For positively charged particles, the lowest Landau
level is occupied by fermions with spin up, and for negatively
charged particles by fermions with spin down. As a result,  each
charged fermion subsystem acquires spin polarization in a magnetic
field.

Further we will consider thermodynamic properties of magnetized
strange quark matter at zero temperature. This approximation is
quite  reasonable  taking into account that for the characteristic
densities in the interior of a neutron star of about several times
nuclear saturation density the temperature is much less than the
quark chemical potentials. Concerning electrons, although finite
temperature effects can be important for them, their contribution to
the thermodynamic quantities is usually unessential because the
electron fraction, in turn, is small. In the zero temperature case,
the thermodynamic potential for an ideal gas of relativistic
fermions of $i$th species in the external magnetic field
reads~\cite{C}
\begin{align}\Omega_i&=-\frac{|q_i|g_i H}{4\pi^2}\sum_{\nu=0}^{\nu_{\mathrm{max}}^i}
(2-\delta_{\nu,0}) \label{omegai0}\\ &\quad \times \biggl\{\mu_i
k^i_{F,\nu}-\bar m_{i,\nu}^2\ln\biggl|\frac{k^i_{F,\nu}+\mu_i}{\bar
m_{i,\nu}}\biggr|\biggr\},\nonumber \end{align}
 where the factor $(2-\delta_{\nu,0})$ takes into account the spin degeneracy of Landau
 levels,
$g_i$ is the remaining degeneracy factor [$g_f=3$ for quarks (number
of colors), and $g_e=1$ for electrons], $\mu_i$ is the chemical
potential, and
\begin{align}\bar
m_{i,\nu}=\sqrt{m_i^2+2\nu|q_i|H},\quad
k^i_{F,\nu}=\sqrt{\mu_i^2-\bar m_{i,\nu}^2}.
\end{align}
In Eq.~\p{omegai0}, summation runs up to
$$\nu_{\mathrm{max}}^i=I\bigl[\frac{\mu_i^2-m_i^2}{2|q_i|H}\bigr],$$
$I[...]$ being an integer part of the value in the brackets.
 The number density
$\varrho_i=-(\frac{\partial\Omega_i}{\partial\mu_i})_T$ of fermions
of $i$th species  is given by \begin{align}\varrho_i=\frac{|q_i|g_i
H}{2\pi^2}\sum_{\nu=0}^{\nu_{\mathrm{max}}^i}
(2-\delta_{\nu,0})k^i_{F,\nu}. \label{ni0}
\end{align}
 The sum
in Eq.~\p{ni0} can be split into two parts representing  the fermion
number densities with spin up and spin down.  As explained earlier,
the only difference between the two sums  is in the term with
$\nu=0$, corresponding to spin-up fermions if they are positively
charged, and to spin-down fermions, if they are negatively charged.
Then the zero temperature expression for the spin polarization
parameter of the $i$th species subsystem reads:
\begin{align}\Pi_i\equiv\frac{\varrho_i^{\uparrow}-\varrho_i^{\downarrow}}{\varrho_i}=
\frac{q_ig_i H}{2\pi^2\varrho_i}\sqrt{\mu_i^2-m_i^2}.\label{poli0}
\end{align}
 In a strong enough
magnetic field, when only a lowest Landau level is occupied by
fermions of $i$th species, a full polarization occurs with
$|\Pi_i|=1$.

 In order to find the chemical potentials of all fermion
species (and, hence, the corresponding particle number densities),
we will use the following conditions. First,  the conservation of
the total baryon number  is implied:
\begin{align}\frac{1}{3}(\varrho_u+\varrho_d+\varrho_s)=\varrho_B,\label{nb}\end{align}
where $\varrho_B$ is the total baryon number density. Further, quark
matter is considered to be charge neutral:
\begin{align}2\varrho_u-\varrho_d-\varrho_s-3\varrho_{e^-}=0.\label{cnc}\end{align}

Because of the weak processes in the quark core of a neutron
star~\cite{AFO}
\begin{align}d&\rightarrow u+e^-+\bar\nu_e,\quad u+e^-\rightarrow d+\nu_e, \label{wd1}\\
s&\rightarrow u+e^-+\bar\nu_e,\quad u+e^-\rightarrow s+\nu_e,\label{wd2}\\
s&+u\leftrightarrow d+u,
\end{align}
all fermion species are assumed to be in a chemical equilibrium with
the corresponding conditions
\begin{align}\mu_d&=\mu_u+\mu_{e^-},\label{mud}\\
\mu_d&=\mu_s. \label{ds}
\end{align}
Here we suppose that neutrinos and antineutrinos freely escape a
neutron star, and, hence, their chemical potentials are set to zero.
Eqs.~\p{nb}, \p{cnc}, \p{mud} and \p{ds}, with account of
Eq.~\p{ni0}, form the full set of the self-consistency equations for
finding the chemical potentials $\mu_i$ of quarks and electrons.

At zero temperature, the energy density $E_i=\Omega_i+\mu_i
\varrho_i$ for fermions of $i$th species reads
\begin{align}E_i&=\frac{|q_i|g_i H}{4\pi^2}\sum_{\nu=0}^{\nu_{\mathrm{max}}^i}
(2-\delta_{\nu,0}) \label{energyi0}\\
&\quad \times \biggl\{\mu_i k^i_{F,\nu}+\bar
m_{i,\nu}^2\ln\biggl|\frac{k^i_{F,\nu}+\mu_i}{\bar
m_{i,\nu}}\biggr|\biggr\}.\nonumber\end{align}

In the MIT bag model, the total energy density $E$,  longitudinal
$p_l$ and transverse $p_t$ pressures in quark matter  are given
by~\cite{FIKPS}
\begin{align}
E&=\sum_i E_i+\frac{H^2}{8\pi}+B,\label{E}\\
p_{l}&=-\sum_i\Omega_i-\frac{H^2}{8\pi}-B,\label{pl}\\
p_{t}&=-\sum_i\Omega_i-HM+\frac{H^2}{8\pi}-B, \label{pt}
\end{align} where $B$ is the bag
constant, and $M=\sum_i M_i=-\sum_i
(\frac{\partial\Omega_i}{\partial H})_{\mu_i}$ is the total
magnetization. At zero temperature, the magnetization of $i$th
fermion species reads
\begin{align}
M_i&=\frac{|q_i|g_i}{4\pi^2}\sum_{\nu=0}^{\nu_{\mathrm{max}}^i}
(2-\delta_{\nu,0}) \label{mi0}\\
&\quad \times \biggl\{\mu_i k^i_{F,\nu}-\bigl[\bar
m_{i,\nu}^2+2\nu|q_i|H\bigr]\ln\biggl|\frac{k^i_{F,\nu}+\mu_i}{\bar
m_{i,\nu}}\biggr|\biggr\}.\nonumber\end{align}

It is seen from Eqs.~\p{pl}, \p{pt} that the magnetic field strength
enters differently to the longitudinal and transverse pressures that
reflects the breaking of the $O(3)$ rotational  symmetry in a
magnetic field. In a strong enough magnetic field, the quadratic on
the magnetic field strength term (the Maxwell term) will be
dominating, leading to increasing the transverse pressure and to
decreasing the longitudinal pressure. Hence, there exists a critical
magnetic field $H_c$, at which the longitudinal pressure vanishes,
resulting in the longitudinal instability of strange  quark matter.
Further we will find the critical magnetic field $H_c$ for the
appearance of the longitudinal instability in a  quark system under
the conditions relevant to the cores of  magnetars. Also, we will
determine the threshold magnetic field beyond which the pressure
anisotropy in strange quark matter becomes significant and cannot be
 disregarded anymore.

Note here a principal difference between Eqs.~\p{pl}, \p{pt}, used
in the present study, and analogous equations in Ref.~\cite{FMRO}.
Namely, the pure field contribution (the Maxwell term) was  missed
in the equations of Ref.~\cite{FMRO}, and, by this reason, the
authors of that work wrongly  arrived at the inequality $p_l>p_t$
for the pressures, contrary to the correct result $p_l<p_t$ (for all
relevant magnetic field strengths and densities, the Maxwell term is
larger than  the term linear on magnetization in Eq.~\p{pt}, as was
clarified earlier~\cite{FIKPS} and is confirmed by our
calculations). Eqs.~\p{pl}, \p{pt} were obtained in
Ref.~\cite{FIKPS} using the functional integration technique of a
quantum field theory~\cite{K}. Also, the similar equations (after
disregarding the bag pressure and small higher order terms
containing $M$) were obtained in Refs.~\cite{IY_PLB12,IY_PRC11} for
a degenerate system  of strongly interacting neutrons in an external
magnetic field  by applying a Fermi liquid approach~\cite{AIP,IY3}.
Moreover, the Maxwell term was also accounted for in the stress
tensor of a relativistic magnetized system of free neutrons  in
Ref.~\cite{Kh}, where a similar conclusion about the possibility of
the collapse of a neutron star along the magnetic field (not the
transverse collapse) was reached.

 Note that Eqs.~\p{E}-\p{pt} were obtained in Ref.~\cite{FIKPS} for
 an abstract magnetized one-component  degenerate Fermi gas. All calculations in
that work were done at fixed chemical potential and zero bag
pressure. We consider a scenario which is  relevant in the
astrophysical context of the problem, namely, we study the effects
of the pressure anisotropy explicitly in magnetized three-flavor
quark system (magnetized strange quark matter) under the additional
constraints of the total baryon number conservation, charge
neutrality and chemical equilibrium with respect to various weak
processes occurring in the system. 
The quark species are characterized by their own chemical potentials
and, together with the electron chemical potential, are determined
from the constraints~\p{nb}, \p{cnc}, \p{mud} and \p{ds} at the
given total baryon number density. Under such formulation of the
problem, it is possible to find how the chemical composition of
strange quark matter, spin polarization of various quark flavors and
EoS change with the magnetic field, putting a special emphasis on
the effect of the pressure anisotropy on the above dependences.

\section{Numerical results and discussion}

As was mentioned in Introduction, strange quark matter can be in
absolutely stable state (strange quark stars), or in metastable
state, which can be stabilized by high enough external pressure
(hybrid stars). The valley of the absolute stability in the model
parameter space is determined from the requirement that, at zero
external pressure and temperature, the energy per baryon for strange
quark matter should be less than that for the most stable $^{56}$Fe
nucleus being about 930~MeV. The maximum allowed bag pressure from
the absolute stability window decreases with the magnetic field
strength~\cite{PFIH} and reaches its maximum
$B\simeq90$~MeV/fm$^3$~\cite{FJ} at $H=0$. In turn, the lower bound
on the bag pressure is established from the requirement that, at
zero temperature and pressure, two-flavor quark matter (composed of
$u$ and $d$ quarks) should be less stable with respect to the iron
nucleus $^{56}$Fe, and, hence, the energy per baryon for two-flavor
quark matter should be larger than that for the nucleus
$^{56}$Fe~\cite{FJ,FW,M}. Further, we will be interested in the
astrophysical scenario, in which strange quark matter is formed in
the core of a strongly magnetized neutron star, and, hence, is
metastable at zero pressure. In numerical calculations, we adopt two
values of the bag constant, $B=100$~MeV/fm$^3$ and
$B=120$~MeV/fm$^3$, which are slightly larger than the upper bound
on $B$ from the absolute stability window.
 The core densities corresponding to these bag
pressures are chosen equal to $\varrho_B=3\varrho_0$ and
$\varrho_B=4\varrho_0$, respectively, which are, in principle,
sufficient to produce deconfinement~\cite{FW}
($\varrho_0=0.16\,\mathrm{fm}^{-3}$ being the nuclear saturation
density).   For the current quark masses, we use the same values as
in Refs.~\cite{C,BPC,FMRO,MFP}, i.e., $m_u=m_d=5$~MeV, and
$m_s=150$~MeV.

\begin{figure}[tb]
\begin{center}
\includegraphics[width=8.6cm,keepaspectratio]{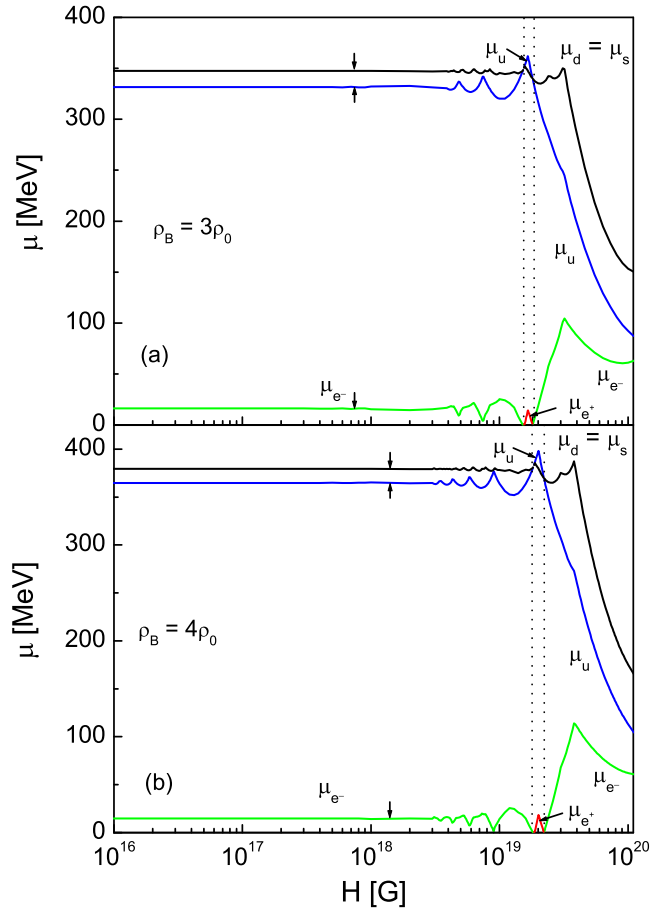}
\end{center}
\vspace{-2ex} \caption{(Color online) Various fermion species
chemical potentials as functions of the magnetic field strength at
zero temperature for the total baryon density (a)
$\varrho_B=3\varrho_0$ and (b) $\varrho_B=4\varrho_0$. The chemical
potential of positrons  is shown by the red curves between the
vertical dotted lines. The vertical arrows  indicate the points
corresponding to the critical field $H_{c}$; see further details  in
the text.} \label{fig1}\vspace{-0ex}
\end{figure}

Fig.~\ref{fig1} shows the chemical potentials of all fermion species
as functions of the magnetic field strength. It is seen that the
chemical potentials of fermions, first, stay practically constant
under increasing the magnetic field, with $d$ and $s$ quark chemical
potentials being somewhat larger (on the value of
$\mu_{e^-}\sim14-16\;\textrm{MeV}$) than the $u$ quark chemical
potential. The apparent Landau oscillations of the chemical
potentials  appear beginning from $H\sim 3\cdot10^{18} -
4\cdot10^{18}$~G, depending on the total baryon number density. At
$H\gtrsim 4\cdot10^{19}$~G, the quark chemical potentials decrease
with the magnetic field. An interesting peculiarity occurs in a
narrow interval near $H\sim 2\cdot10^{19}$~G, marked by the vertical
dotted lines. Namely, for magnetic field strengths from that
interval the $u$ quark chemical potential is larger than the $d$ and
$s$ quark chemical potentials, $\mu_u>\mu_d=\mu_s$. Hence, for such
magnetic fields, according to Eq.~\p{mud}, the electron chemical
potential would be negative, $\mu_{e^-}<0$. If to recall the finite
temperature expression for the electron number density\footnote{With
account of the spin degeneracy factor $(2-\delta_{\nu,0})$, it
reads~\cite{C}:
\begin{align*} \varrho_{e^-}&=\frac{|q_{e}| H}{2\pi^2}\sum_{\nu=0}^\infty
(2-\delta_{\nu,0})\int_0^\infty \,dk_z
\biggl(\frac{1}{e^{\beta(\varepsilon^{e}_\nu-\mu_{e^-})}+1}\\
&\quad- \frac{1}{e^{\beta(\varepsilon^{e}_\nu+\mu_{e^-})}+1}\biggr).
\nonumber \end{align*}},
 its zero temperature limit at $\mu_{e^-}<0$ is, formally, negative,
 contrary to the constraint $\varrho_{e^-}\geqslant0$.
 In fact, this means that  in this interval on $H$  electrons are missing
 and, hence,
the weak $\beta^-$ processes \p{wd1}, \p{wd2} are impossible.
However, for such magnetic fields, the following
 weak $\beta^+$ processes become allowable
\begin{align}u&\rightarrow d+e^++\nu_e,\quad d+e^+\rightarrow u+\bar\nu_e, \label{wd3}\\
u&\rightarrow s+e^++\nu_e,\quad s+e^+\rightarrow
u+\bar\nu_e.\label{wd4}
\end{align}

Hence, for this specific range of the magnetic field strengths, the
charge neutrality and chemical equilibrium conditions should read
\begin{gather}2\varrho_u-\varrho_d-\varrho_s+3\varrho_{e^+}=0,\\
\mu_u=\mu_d+\mu_{e^+},\quad \mu_d=\mu_s,
\end{gather}
which should be solved jointly with the condition of the total
baryon number conservation, Eq.~\p{nb}. The quark  and positron
chemical potentials obtained as solutions of these equations are
shown graphically in Fig.~\ref{fig1} as the corresponding curves
between the vertical dotted lines. With increasing the core density,
the width of the interval on $H$, where positrons appear,  increases
slightly as well (cf. the ranges
$1.56\cdot10^{19}$\,G\,--\,$1.80\cdot10^{19}$\,G at
$\varrho_B=3\varrho_0$ and
$1.86\cdot10^{19}$\,G\,--\,$2.21\cdot10^{19}$\,G at
$\varrho_B=4\varrho_0$). Thus, as a matter of principle, in strongly
magnetized strange quark matter at zero temperature, subject to the
total baryon number conservation, charge neutrality and chemical
equilibrium conditions, positrons can appear in a certain narrow
interval of the magnetic field strengths, replacing electrons. In
this case, strange quark matter will have negative hadronic electric
charge.
Note that, according to Ref.~\cite{FJ}, the contact of stable
strange quark matter, having negative hadronic electric charge, with
the ordinary matter would have the disastrous consequences for the
latter, because positively charged nuclei would be attracted to
strange quark matter and absorbed. However, the contact of
metastable strange quark matter, having negative hadronic electric
charge, with hadronic matter in the interior of a neutron star is
possible, because  the outer hadronic layer provides the  necessary
  external pressure to stabilize strange quark
matter in the core and cannot be completely depleted. Nevertheless,
we should calculate the critical field $H_{c}$ for the appearance of
the longitudinal instability, which could prevent the occurrence of
positrons in a certain range of magnetic field strengths with
$H\gtrsim10^{19}$~G. The meaning of the vertical arrows in
Fig.~\ref{fig1} will be discussed later in the text.

\begin{figure}[tb]
\begin{center}
\includegraphics[width=8.6cm,keepaspectratio]{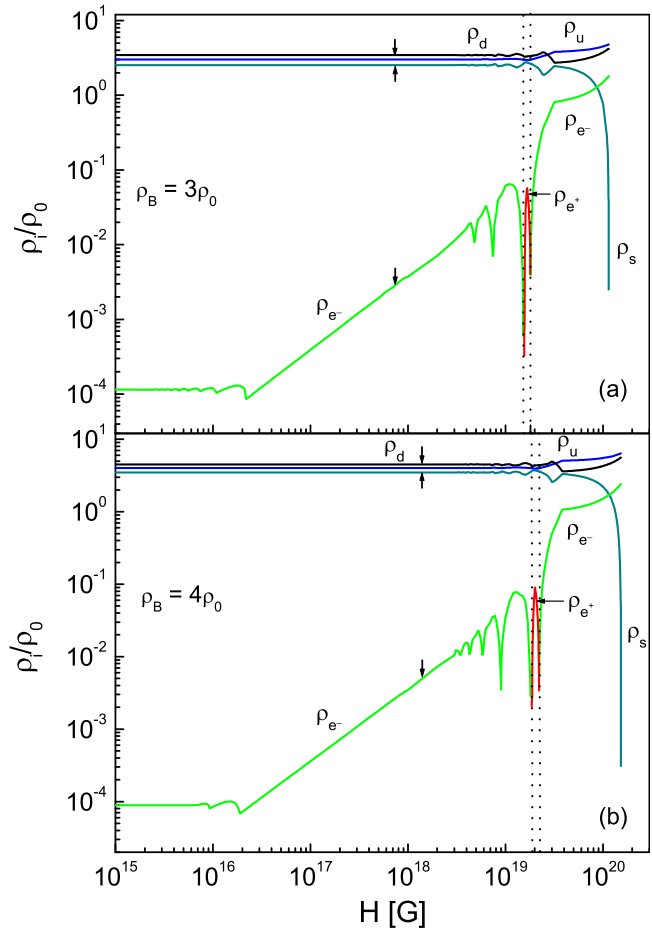}
\end{center}
\vspace{-2ex} \caption{(Color online) Same as in Fig.~\ref{fig1} but
for the particle number densities $\varrho_i/\varrho_0$ of various
fermion species.  The particle number density of positrons  is shown
by the red curves between the vertical dotted lines. }
\label{fig2}\vspace{-0ex}
\end{figure}

Fig.~\ref{fig2} shows the abundances of various fermion species as
functions of the magnetic field strength. The number densities of
$u$ and $d$ quarks are quite close  to each other for all magnetic
fields under consideration. The electron number density begins quite
rapidly to increase at $H\approx 2.2\cdot10^{16}$~G for
$\varrho_B=3\varrho_0$ and at $H\approx 1.9\cdot10^{16}$~G for
$\varrho_B=4\varrho_0$. As noted earlier, in the narrow interval
near $H\sim 2\cdot10^{19}$~G electrons are replaced by positrons,
and beyond this interval electrons appear again with the number
density increasing with $H$. The $s$ quark content of strange quark
matter stays practically constant till the field strength
$H\approx4.1\cdot10^{18}$~G at $\varrho_B=3\varrho_0$ and
$H\approx3.8\cdot10^{18}$~G at $\varrho_B=4\varrho_0$, beyond which
 the $s$ quark number density experiences visible Landau
oscillations. Then,  beginning from the field strength
$H\approx3.2\cdot10^{19}$~G at $\varrho_B=3\varrho_0$ and
$H\approx3.9\cdot10^{19}$~G at $\varrho_B=4\varrho_0$,   the $s$
quark content rapidly decreases. Strange quark matter loses its
strangeness and  turns into two-flavor quark matter in the magnetic
fields slightly larger than $10^{20}$~G. Again, we should determine
the critical field $H_{c}$ in order to check whether this
significant drop of strangeness could happen in a strong magnetic
field.

\begin{figure}[tb]
\begin{center}
\includegraphics[width=8.6cm,keepaspectratio]{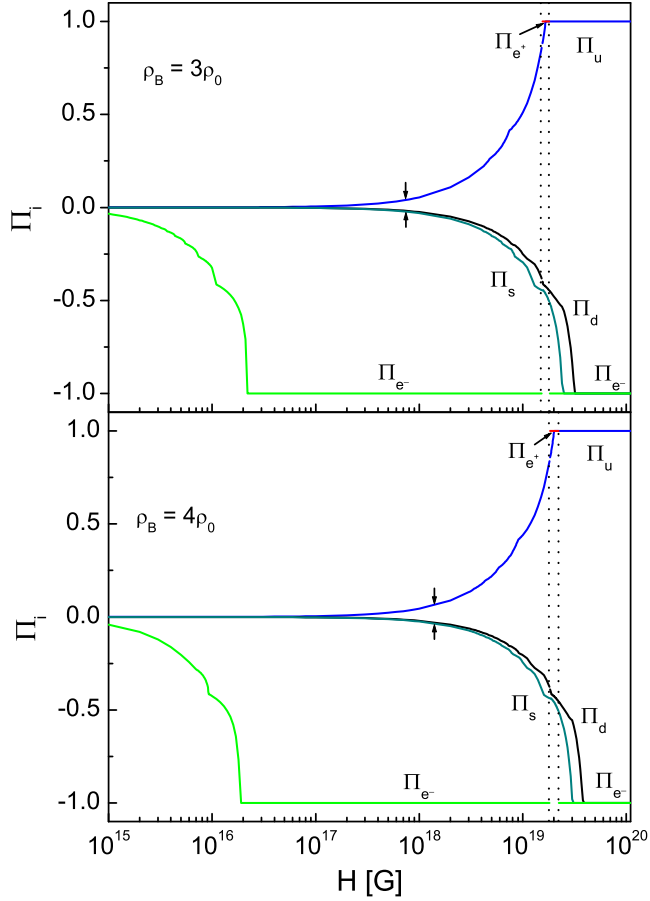}
\end{center}
\vspace{-2ex} \caption{(Color online) Same as in Fig.~\ref{fig1} but
for the spin polarization parameter
 of various fermion
species. The spin polarization parameter
 of positrons  is shown by the red segments
between the vertical dotted lines. } \label{fig3}\vspace{-0ex}
\end{figure}

Fig.~\ref{fig3} shows the spin polarization parameter $\Pi_i$ for
various fermion species, determined according to Eq.~\p{poli0}, as a
function of the magnetic field strength. Spin polarization of $u$
quarks is positive while for $d,s$ quarks and electrons it is
negative. The magnitude of the spin polarization parameter $\Pi_i$
increases with $H$ till it is saturated at the respective saturation
field $H_s^i$. At $H=H_s^i$, the corresponding $i$th fermion species
becomes fully spin polarized. The respective values of the
saturation field are: $H_{s}^{e}\approx2.2\cdot10^{16}$~G for
electrons, $H_{s}^u\approx1.7\cdot10^{19}$~G for $u$ quarks,
$H_{s}^s\approx2.5\cdot10^{19}$~G for $s$ quarks and
$H_{s}^d\approx3.2\cdot10^{19}$~G for $d$ quarks at
$\varrho_B=3\varrho_0$, and $H_{s}^{e}\approx1.9\cdot10^{16}$~G for
electrons, $H_{s}^u\approx2.0\cdot10^{19}$~G for $u$ quarks,
$H_{s}^s\approx3.1\cdot10^{19}$~G for $s$ quarks and
$H_{s}^d\approx3.9\cdot10^{19}$~G for $d$ quarks at
$\varrho_B=4\varrho_0$. Note that quite a rapid increase of the
electron number density with the magnetic field (cf.
Fig.~\ref{fig2}) begins just at the saturation field $H_s^e$,   and,
hence, this increase occurs when electrons become completely spin
polarized. Further oscillations in the electron number density are,
in fact, caused by the Landau oscillations of the quark number
densities, which influence the electron population through the
charge neutrality condition. The significant drop of the strange
quark content begins just at the magnetic field strength at which
the $d$ quarks become completely spin polarized. Although the
$s$-quark current mass is larger than that for $d$ quark, $m_s>m_d$,
$s$~quarks become fully polarized at a smaller saturation field
because their particle density is smaller than  for $d$~quarks,
$\varrho_s<\varrho_d$. The spin polarization parameter of various
fermion species in the magnetic field range, where positrons appear,
is shown by the respective curves between the vertical dotted lines.
It is seen that positrons occur already fully polarized, and $u$
quarks become totally polarized just in this range of the magnetic
field strengths. Nevertheless, as mentioned before, only after
determining the critical field $H_c$ for the appearance of the
longitudinal instability, it would be possible to determine the
degree of spin polarization which could be reached for each of the
fermion species.

\begin{figure}[tb]
\begin{center}
\includegraphics[width=8.6cm,keepaspectratio]{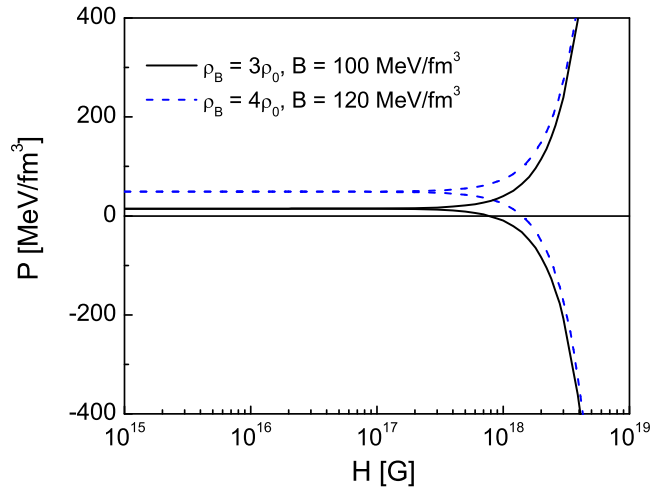}
\end{center}
\vspace{-2ex} \caption{(Color online)  Transverse  (ascending
branches in the branching curves) and longitudinal (descending
branches in the branching curves) pressures in magnetized strange
quark matter at zero temperature as functions of the magnetic field
strength for $\varrho_B=3\varrho_0, B=100\,\mathrm{MeV/fm^3}$ (solid
lines) and $\varrho_B=4\varrho_0, B=120\,\mathrm{MeV/fm^3}$ (dashed
lines).} \label{fig5}\vspace{-0ex}
\end{figure}

Now we present the results of calculations of the longitudinal $p_l$
and transverse $p_t$ pressures. Fig.~\ref{fig5} presents these
quantities as functions of the magnetic field strength at
$\varrho_B=3\varrho_0, B=100\,\mathrm{MeV/fm^3}$ (solid lines) and
$\varrho_B=4\varrho_0, B=120\,\mathrm{MeV/fm^3}$ (dashed lines). It
is seen that the transverse (ascending branches) and longitudinal
(descending branches) pressures, first, stay practically constant
and indistinguishable from each other. This behavior of $p_t$ and
$p_l$ corresponds to the isotropic regime. Beyond some threshold
magnetic field $H_{th}$,  the transverse pressure $p_t$ increases
with $H$ while the longitudinal pressure $p_l$ decreases with it,
clearly reflecting the anisotropic nature of the total pressure in
strange quark matter in such strong magnetic fields  (anisotropic
regime). In the critical magnetic field $H_c$, the longitudinal
pressure $p_l$ vanishes. This happens at
$H_{c}\approx7.4\cdot10^{17}$~G for $\varrho_B=3\varrho_0,
B=100\,\mathrm{MeV/fm^3}$, and at $H_{c}\approx1.4\cdot10^{18}$~G
for $\varrho_B=4\varrho_0, B=120\,\mathrm{MeV/fm^3}$. Above the
critical magnetic field, the longitudinal pressure is negative
leading to the longitudinal instability of strange quark matter.
Therefore, the thermodynamic properties of strange quark matter
should be considered in the magnetic fields $H<H_c$.  In fact, the
critical field sets the upper bound on the magnetic field strength
which can be reached in the core of a strongly magnetized hybrid
star. For comparison, we  present also the values of the critical
field for dense neutron matter with the Skyrme BSk20 interaction at
zero temperature being $H_c\approx 1.6 \cdot 10^{18}$~G  at
$\varrho_B = 3\varrho_0$, and $H_c \approx 2.4 \cdot 10^{18}$~G  at
$\varrho_B = 4\varrho_0$~\cite{IY_PRC11}.

Now, in order to see, which of the discussed already  features of
strange quark matter at zero temperature in a strong magnetic field
are preserved before the appearance of the longitudinal instability,
we show in Figs.~\ref{fig1}-\ref{fig3} by the vertical arrows the
respective values of  the physical quantities corresponding to the
critical field $H_c$. Let us begin with Fig.~\ref{fig1} for the
chemical potentials of various fermion species. It is seen that the
chemical potentials of quarks and electrons stay practically
unchanged before the appearance of the longitudinal instability. The
significant changes in the chemical potentials occur only in the
fields $H>H_c$.  In particular, the longitudinal instability
precludes the appearance of positrons for which the fields
$H\gtrsim10^{19}$~G are necessary.

Let us turn to Fig.~\ref{fig2} for the abundances of various fermion
species. Till the  critical field $H_c$, the content of quark
species stays practically constant while the electron fraction
remains quite small, $\varrho_{e^-}/\varrho_0\lesssim10^{-2}$. Also,
there is no room for the significant drop of the strange quark
content in  strong magnetic fields $H\sim10^{20}$~G which is averted
by the appearance of the longitudinal instability in the critical
field $H_c$. In fact, despite the presence of strong magnetic fields
$H\sim10^{18}$~G, strange quark matter has the same fraction of $s$
quarks as in the field-free case.

Let us now consider Fig.~\ref{fig3} for spin polarizations of
various fermion species. It is seen  that the full polarization in a
strong magnetic field can be achieved only for electrons. For
various quark species, the spin polarization remains quite moderate
up to the critical magnetic field $H_c$. E.g., at
$\varrho_B=4\varrho_0, H=H_c$ we have $\Pi_u\approx0.06$,
$\Pi_d\approx-0.03$, $\Pi_s\approx-0.04$; at
$\varrho_B=3\varrho_0,H=H_c$, the quark spin polarizations are
similar to these values with the maximum magnitude of the spin
polarization parameter for $u$ quarks, $\Pi_u\approx0.04$.
Therefore, the occurrence of a field-induced fully polarized state
in strange quark matter is prevented by the appearance of the
longitudinal instability in the critical magnetic field. The degree
of spin polarization of various constituents is an important issue
for determining the neutrino mean free paths in magnetized strange
quark matter \cite{SSB}, and, hence, it is relevant for the adequate
description of the neutrino transport and thermal evolution of a
magnetized  hybrid star.

\begin{figure}[tb]
\begin{center}
\includegraphics[width=8.6cm,keepaspectratio]{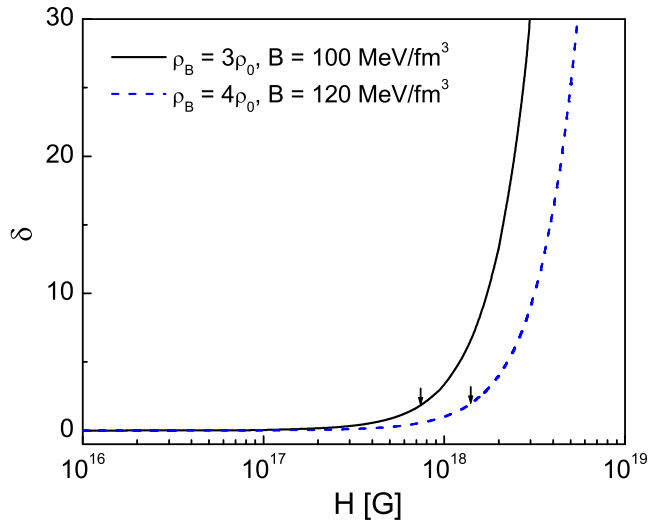}
\end{center}
\vspace{-2ex} \caption{(Color online) Same as in Fig.~\ref{fig5} but
for the normalized difference $\delta=\frac{p_t-p_l}{p_0}$ between
the transverse and longitudinal pressures. The vertical arrows show
the maximum normalized splitting $\delta_c$ at the corresponding
critical field $H_c$.} \label{fig6}\vspace{-0ex}
\end{figure}

Thus, in the anisotropic regime, the pressure anisotropy in a strong
magnetic field plays an important role and should be accounted for
in the description of the thermodynamic properties  of strange quark
matter. Let us now make the estimate of the threshold magnetic field
$H_{th}$ above which the pressure anisotropy cannot be disregarded.
Fig.~\ref{fig6} shows the normalized difference between the
transverse and longitudinal pressures
$$\delta=\frac{p_t-p_l}{p_0},$$
where $p_0$ is the isotropic pressure (which corresponds to the weak
field limit with $p_l=p_t=p_0$), as a function of the magnetic field
strength for the cases under consideration. Following
Refs.~\cite{FIKPS,IY_PLB12,IY_PRC11}, for finding the threshold
field $H_{th}$ one can use the approximate criterion
$\delta\simeq1$. Then anisotropic regime enters at
$H_{th}\approx5.5\cdot10^{17}$~G for $\varrho_B=3\varrho_0,
B=100\,\mathrm{MeV/fm^3}$, and at $H_{th}\approx9.9\cdot10^{17}$~G
for $\varrho_B=4\varrho_0, B=120\,\mathrm{MeV/fm^3}$.  For
comparison~\cite{IY_PRC11}, the threshold field for neutron matter
with the BSk20 Skyrme interaction at zero temperature is
$H_{th}\approx 1.2 \cdot10^{18}$~G for $\varrho_B = 3\varrho_0$, and
$H_{th}\approx 1.8 \cdot10^{18}$~G for $\varrho_B = 4\varrho_0$. The
anisotropy parameter $\delta$ reaches its maximum $\delta_c\sim2$ in
the critical field $H_c$, corresponding to the onset of the
longitudinal instability in strange quark matter.  In the
anisotropic regime, a hybrid star is deformed and takes the oblate
form. Thus, as follows from the previous discussions, the effects of
the pressure anisotropy are important at $H_{th}< H<H_{c}$, and
significantly influence the structural  and polarization properties
of the quark core in  a strongly magnetized neutron star.

\begin{figure}[tb]
\begin{center}
\includegraphics[width=8.6cm,keepaspectratio]{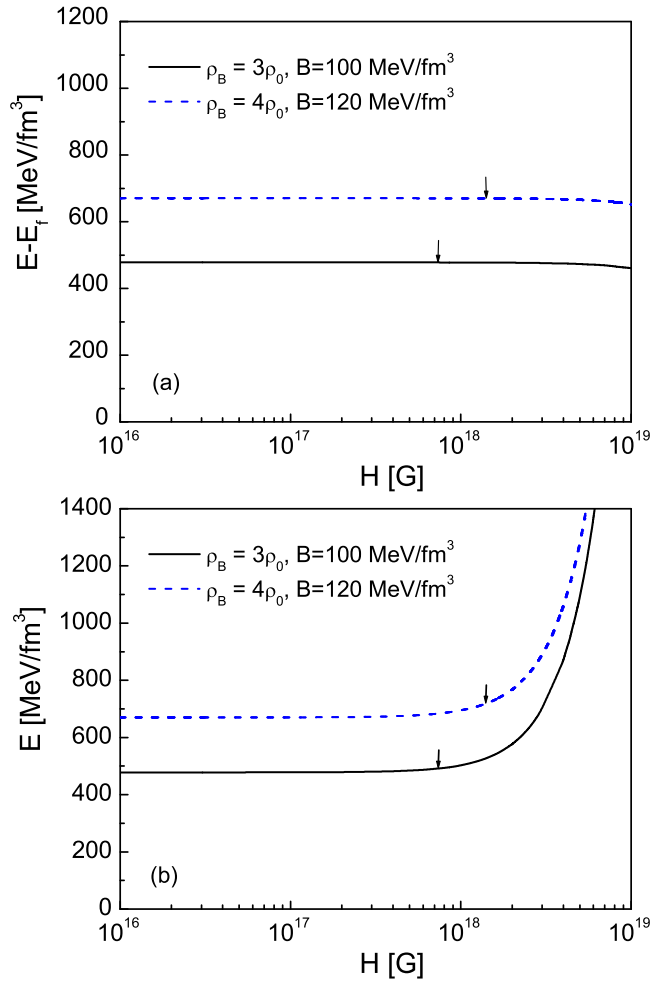}
\end{center}
\vspace{-2ex} \caption{(Color online) Same as in Fig.~\ref{fig5} but
for the energy density $E$ of the system (a) without the  magnetic
field energy density contribution $E_f=\frac{H^2}{8\pi}$ and (b)
with account of it. The vertical arrows indicate the points
corresponding to the
 critical field $H_c$.} \label{fig7}\vspace{-0ex}
\end{figure}

Fig.~\ref{fig7} shows the energy density $E$ of the system without
the magnetic field energy density contribution
$E_f=\frac{H^2}{8\pi}$ (top panel) and with account of it (bottom
panel) as a function of the magnetic field strength at zero
temperature. It is seen that, due to the Landau diamagnetism, the
energy density of solely magnetized strange quark matter decreases
with the magnetic field. However, the overall effect of the magnetic
field, with account of the pure magnetic field contribution $E_f$,
is to increase the energy density of the system. Nevertheless, this
effect of the magnetic field is, in fact, insignificant because the
magnetic field is bound from above by the critical magnetic field
$H_c$. The values of the energy density $E$, corresponding to the
critical field $H_c$, are shown in Fig.~\ref{fig7} by the vertical
arrows.

\begin{figure}[tb]
\begin{center}
\includegraphics[width=8.6cm,keepaspectratio]{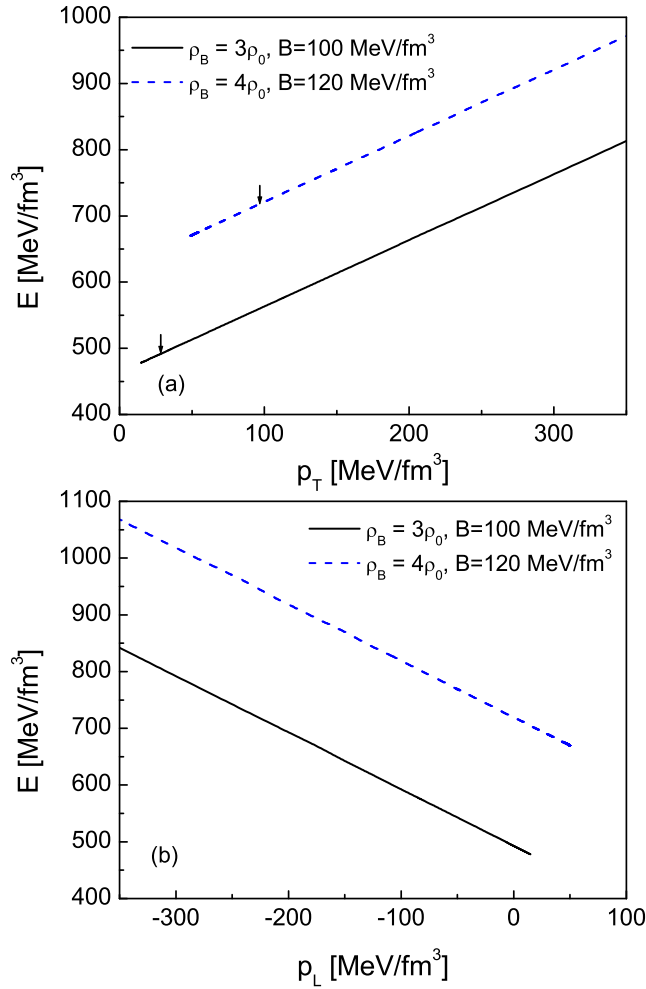}
\end{center}
\vspace{-2ex} \caption{(Color online) The energy density $E$ of the
system at zero temperature
 as a function of: (a) the transverse pressure $p_t$ and (b) the longitudinal pressure
  for the
cases $\varrho_B=3\varrho_0, B=100\,\mathrm{MeV/fm^3}$ and
$\varrho_B=4\varrho_0, B=120\,\mathrm{MeV/fm^3}$. The meaning of the
vertical arrows in the top panel is the same as in Fig.~\ref{fig7}.
In the bottom panel, the physical region corresponds to $p_l>0$.}
\label{fig8}\vspace{-0ex}
\end{figure}

Because of the pressure anisotropy, the equation of state of strange
quark matter in a strong magnetic field is also anisotropic.
Fig.~\ref{fig8} shows the dependence of the energy density $E$ of
the system on the transverse  pressure $p_t$ (top panel) and on the
longitudinal pressure $p_l$ (bottom panel) after excluding the
dependence on $H$ in these quantities. In particular, the
anisotropic character  of the pressure is reflected in the fact that
the energy density is the increasing function of $p_t$ while it
decreases with $p_l$. This is because the dominant Maxwell term
enters the transverse pressure $p_t$ and the energy density $E$ with
positive sign while it enters the longitudinal pressure $p_l$ with
negative sign.  The vertical arrows in the top panel indicate the
points in these lines corresponding to the critical field $H_c$. In
the bottom panel, the physical region corresponds to $p_l>0$.

Note that because the EoS of strange quark matter becomes
essentially anisotropic in an ultrastrong magnetic field, the usual
scheme for finding the mass-radius relationship based on the
Tolman-Oppenheimer-Volkoff (TOV) equations~\cite{TOV} for a
spherically symmetric and static compact star should be revised.
Instead, the corresponding relationship should be found by the
self-consistent treatment of the anisotropic EoS and axisymmetric
TOV equations substituting the conventional TOV equations in the
case of an axisymmetric compact star.

In summary, we have considered the impact of  strong magnetic fields
up to $10^{20}$~G on the thermodynamic properties of strange quark
matter at zero temperature under additional constraints of total
baryon number conservation, charge neutrality and chemical
equilibrium with respect to various weak processes occurring in the
system. The study has been done within the framework of the MIT bag
model with the finite current quark masses $m_u=m_d\not=m_s$.    In
the numerical calculations, we have adopted two sets of the total
baryon number density and bag pressure, $\varrho_B=3\varrho_0,
B=100\,\mathrm{MeV/fm^3}$ and $\varrho_B=4\varrho_0,
B=120\,\mathrm{MeV/fm^3}$. It has been found that in strong magnetic
fields up to $10^{20}$~G some interesting features in the chemical
composition and spin structure of strange quark matter could occur:

(1) The content of strange quarks rapidly decreases  in the fields
somewhat larger than $10^{19}$~G and becomes negligible in the
fields slightly exceeding $10^{20}$~G;

(2) For the magnetic field strengths in the quite narrow interval
near $H\sim2\cdot10^{19}$~G the constraints of total baryon number
conservation, charge neutrality and chemical equilibrium can be
satisfied only if positrons appear in various weak processes in that
range of the field strengths (instead of electrons);

(3) Electrons occupy only the lowest Landau level and, hence, become
completely spin polarized in the magnetic fields somewhat larger
than $10^{16}$~G; $u$, $s$ and $d$ quarks  become fully polarized in
the fields somewhat larger than $10^{19}$~G (the recitation of the
quark species  is in the order in which they appear fully polarized
 under increasing $H$).

Nevertheless, under such strong magnetic fields, the total pressure
containing also the magnetic field contribution, becomes
anisotropic, and the effects of the pressure anisotropy change most
of the above conclusions. Namely, the longitudinal (along the
magnetic field) pressure decreases with the magnetic field (contrary
to the transverse pressure increasing with $H$) and vanishes in the
critical field $H_c$ resulting in the longitudinal instability of
strange quark matter. The value of the critical field $H_c$ depends
on the total baryon number density of strange quark matter and the
bag pressure $B$, and it turns out to be somewhat less or larger
than $10^{18}$~G for the two sets of the parameters,
 considered in the
given study.  Therefore, the appearance of the longitudinal
instability in strong magnetic fields beyond the critical one
precludes the features (1), (2) in the chemical composition of
strongly magnetized strange quark matter. Concerning the conclusion
(3), only electrons can reach the state of  full polarization, that
is not true for quarks of all flavors, whose polarization remains
mild even for magnetic fields near $H_c$.

The pressure anisotropy becomes relevant beyond some threshold field
$H_{th}$. For the sets of the
 total baryon number density and bag
constant considered in the given study, it turns out  that
$10^{17}<H_{th}\lesssim10^{18}$~G.  This estimate is somewhat less
than that found for strongly magnetized dense neutron matter,
$H_{th}\sim10^{18}$~G~\cite{IY_PLB12,IY_PRC11}. In strong magnetic
fields $H>H_{th}$, the EoS of strange quark matter becomes
essentially anisotropic. The longitudinal and transverse pressures
as well as the anisotropic EoS of magnetized strange quark matter
have been determined at the total baryon number densities and
magnetic field strengths relevant to the interiors of magnetars.

In this work, we have studied the impact of a strong magnetic field
on the thermodynamic properties of strange quark matter at zero
temperature. It would be also of interest to extend this research to
finite temperatures~\cite{IY_PRC11,DMS}, which can lead to a number
of interesting effects, such as, e.g, an unusual behavior of the
entropy of a spin polarized state~\cite{IY2,I07}.

In conclusion, it is worthy to note that, because strong magnetic
fields of about $H\sim10^{18}$~G (RHIC), or even by order of
magnitude larger (LHC), are generated in non-central high-energy
heavy-ion collisions \cite{SIT,DH}, the effects of the pressure
anisotropy should be relevant there as well. In particular, the
pressure anisotropy in a strong magnetic field can contribute to the
enhancement of the elliptic flow of hot nuclear matter created in a
heavy-ion collision~\cite{T}. Because the conditions in high-energy
heavy-ion collisions are different from those in the cores of
strongly magnetized neutron stars (the absence of chemical
equilibrium with respect to the weak processes, the low baryon
number densities and high temperatures), the possibility for the
occurrence of the strangeness suppression in a strong magnetic field
for the former case needs a separate study.

J.Y. was supported by grant 2010-0011378 from Basic Science Research
Program through NRF of Korea funded by MEST and by grant R32-10130
from WCU project of MEST and NRF.

\end{document}